\documentclass{aip-cp}

\usepackage[numbers]{natbib}
\usepackage{rotating}
\usepackage{graphicx}

\graphicspath{{./figs/}}

\begin{document}

\title{Observables of QCD Diffraction}

\author[aff1]{Mikael Mieskolainen\corref{cor1}}
\author[aff1,aff2]{Risto Orava}
\eaddress{risto.orava@cern.ch}

\affil[aff1]{Department of Physics, University of Helsinki and Helsinki Institute of Physics,
P.O. Box 64, FI-00014, Finland}
\affil[aff2]{CERN, CH-1211 Geneva 23, Switzerland}
\corresp[cor1]{Corresponding author: mikael.mieskolainen@cern.ch}

\maketitle

\begin{abstract}
A new combinatorial vector space measurement model is introduced for soft QCD diffraction. The model independent mathematical construction resolves experimental complications; the theoretical framework of the approach includes the Good-Walker view of diffraction, Regge phenomenology together with AGK cutting rules and random fluctuations.
\end{abstract}

\section{INTRODUCTION}

Soft diffraction bases theoretically on (soft) Pomeron exchange, the vacuum singularity of Regge theory. In QCD, this is described as a non-local (long-wavelength) gluonic color singlet ladder based object. However, a complete theoretical description of the multi-Pomeron exchanges and interactions, the exact nature of Reggeon exchanges and random fluctuations in hadronization process are still lacking. All these processes can generate large rapidity gaps (LRGs) of several units. An alternative approach is to treat high energy hadronic diffraction as a coherent process, where the ”relativistic wave function”, and its components, undergo unitary scatterings and absorptions. This well known Good-Walker picture \cite{good1960diffraction} is usually implemented by using multichannel eikonal models to account for the complicated proton structure and its coherent fluctuations \cite{kaidalov2001probabilities}, albeit in an integrated way.

Experimentally, soft diffraction is traditionally equated with registering large rapidity gap events. In practice, a number of approximations are required for defining and measuring the rapidity gaps as pseudorapidity intervals void of particles. There are several limitations in defining the rapidity gaps as experimental observables. First of all, electrically neutral particles often remain unaccounted for, even if their presence can be partially inferred using the measured secondaries. Low mass diffractive systems require very forward instrumentation which, in general, covers only part of the small angle scattering at the LHC. Theoretically and experimentally, the low mass resonance region of single and double diffraction, and high mass asymptotic are poorly understood. Counting events with widely varying definitions of rapidity gaps and relatively large $E_\perp/p_\perp$ thresholds cannot give a complete view on the subject of QCD diffraction.

In the following, binary vector spaces over the number field $\mathbb{F}_2=\{0,1\}$ are defined for the chosen observables, then a probabilistic extraction of diffractive cross sections and, finally, new combinatorial approaches are introduced for probing issues like the famous AGK cutting rules \cite{abramowsky1974agk}. In all this, the experimental limitations together with the above theoretical motivations are used as a guidance.

\section{DEFINING THE OBSERVABLES}

\begin{figure}[ht]
\includegraphics[width=150pt]{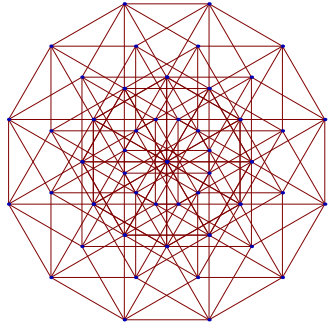}
  \caption{Binary vector space as a $d$-hypercube graph or lattice. Each vertex represents one vector in $\mathbb{F}_2^d$, here $d = 6$. These vertices and weights associated with them are interpreted directly as different final state configurations, corresponding to the partial cross sections $\sigma_k$.}
  \label{fig: d6}
\end{figure}

Now instead of merely counting rapidity gaps, the partial cross sections $\sigma_{inel}^{pp} \equiv \sigma_1 + \sigma_2 + \sigma_3 + \dots + \sigma_n$ are considered, where each $\sigma_k$ corresponds to one particular final state "topology class" over a finite $d$ interval discretized pseudorapidity axis, integrated over the transverse $\varphi$-plane. These span $k = 1, \dots, 2^d -1 = n$ non-zero binary vectors in $\mathbb{F}_2^d$, where each vector component Bernoulli random variable $X_i \in \{0,1\}$, $1 \leq i \leq d$, encodes whether at least one final state in the given rapidity (detector) slice is observed. Simplified, these components correspond to pseudorapidity intervals of $[\eta_i^{\textrm{min}}, \eta_i^{\textrm{max}}]$ which represent geometric projection boundaries. Experimentally, these intervals may overlap or remain distinct from each others. In the limit $d \rightarrow \infty$ the problem turns into track counting, and at $d = 1$, any sensitivity to different process cross sections is lost. Depending on the exact definition, the number of distinct "topology classes" is not necessarily $n$, the number of binary combinations. The corresponding abstract final state vector space $\mathbb{F}_2^d$ is illustrated in Figure \ref{fig: d6}.
\vspace{-0.6em}

Events with varying rapidity gap combinations are contained in this space. In practice, due to finite statistics and finite discretization, there is both Poissonian and discretization uncertainty influence the exact gap sizes. It is possible to use also a different number field, such as the real valued vector space $\mathbb{R}^d$ as in \cite{mieskolainen2014bayesian}, where it was shown that large rapidity gaps are obtained as a particular limit of the multivariate space. In principle, this allows event-by-event utilization of $p_\perp$ or multiplicity degrees of freedom. Real valued distributions can be constructed also "after" the binary subspace projection, as a hybrid approach. The main benefit in using the binary vector spaces, is to allow concise algebraic representations of the measurement itself, and to factorize the model parameter extractions in addition.

Visible or fiducial partial cross sections $\sigma_k^{(vis)}$ are defined in terms of $(\eta,p_\perp)$ acceptance and efficiency functions of charged and (or) neutral particles for each $i$-th discretized pseudorapidity interval. Crucial experimental issue is the separation of efficiency corrections on visible part versus the pure extrapolation to outside of the fiducial acceptance region. This is difficult in the forward domain, due to limited granularity of the calorimetry, tracking and intense fluxes of secondaries from the interactions in the beam pipe and surrounding material. Low $p_\perp$-thresholds and high-$|\eta|$ coverage are of utmost importance. Matrix unfolding procedure is necessary in order to turn the visible detector level partial cross sections to the particle level cross sections $U: \{\sigma_k^{(vis)}\} \mapsto \{\sigma_k\}$. Detector inefficiencies tend to create artificial rapidity gaps and this is to be taken into account in the unfolding process.

\begin{figure}[h]
\includegraphics[width=160pt]{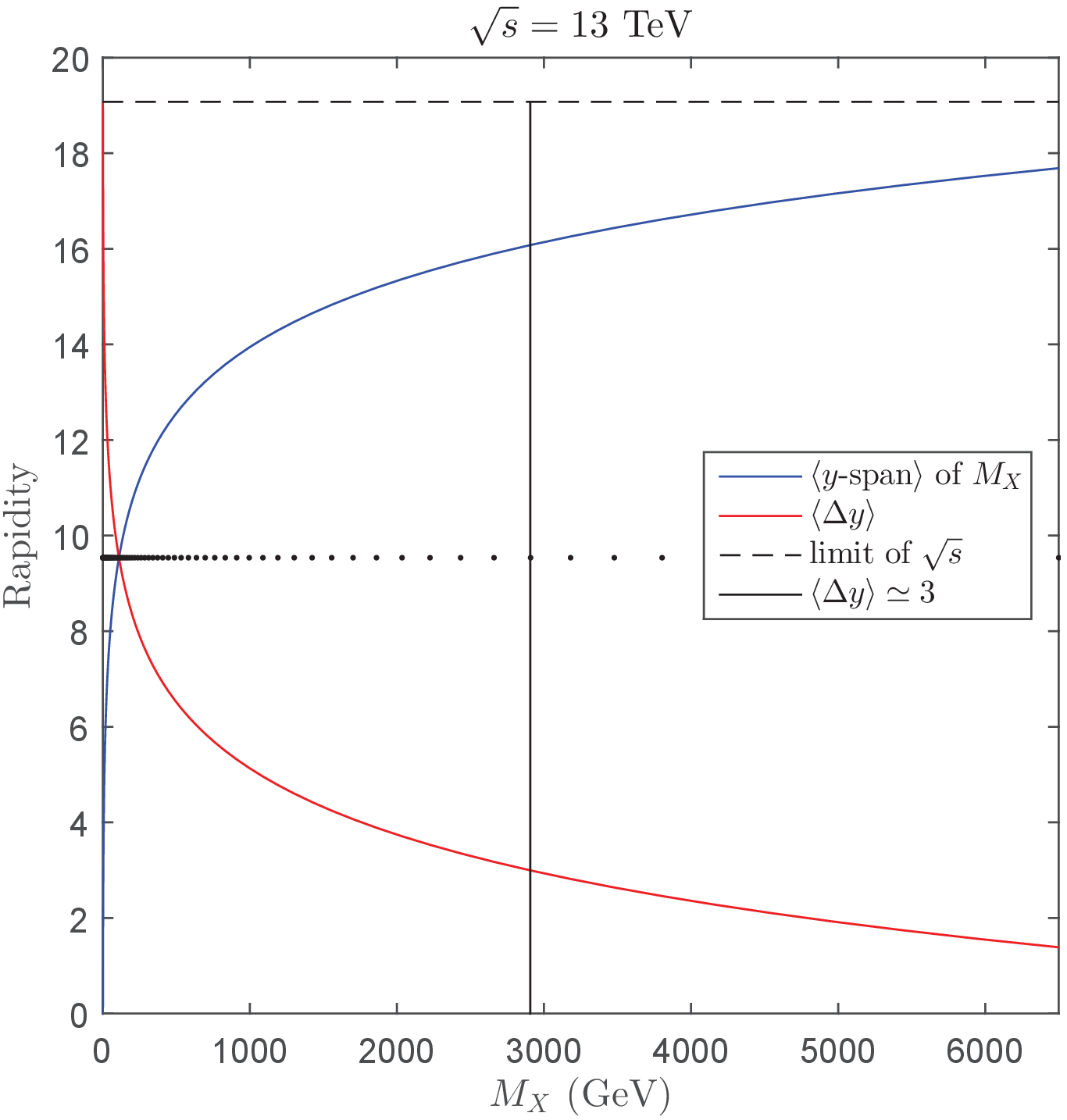}
\includegraphics[width=160pt]{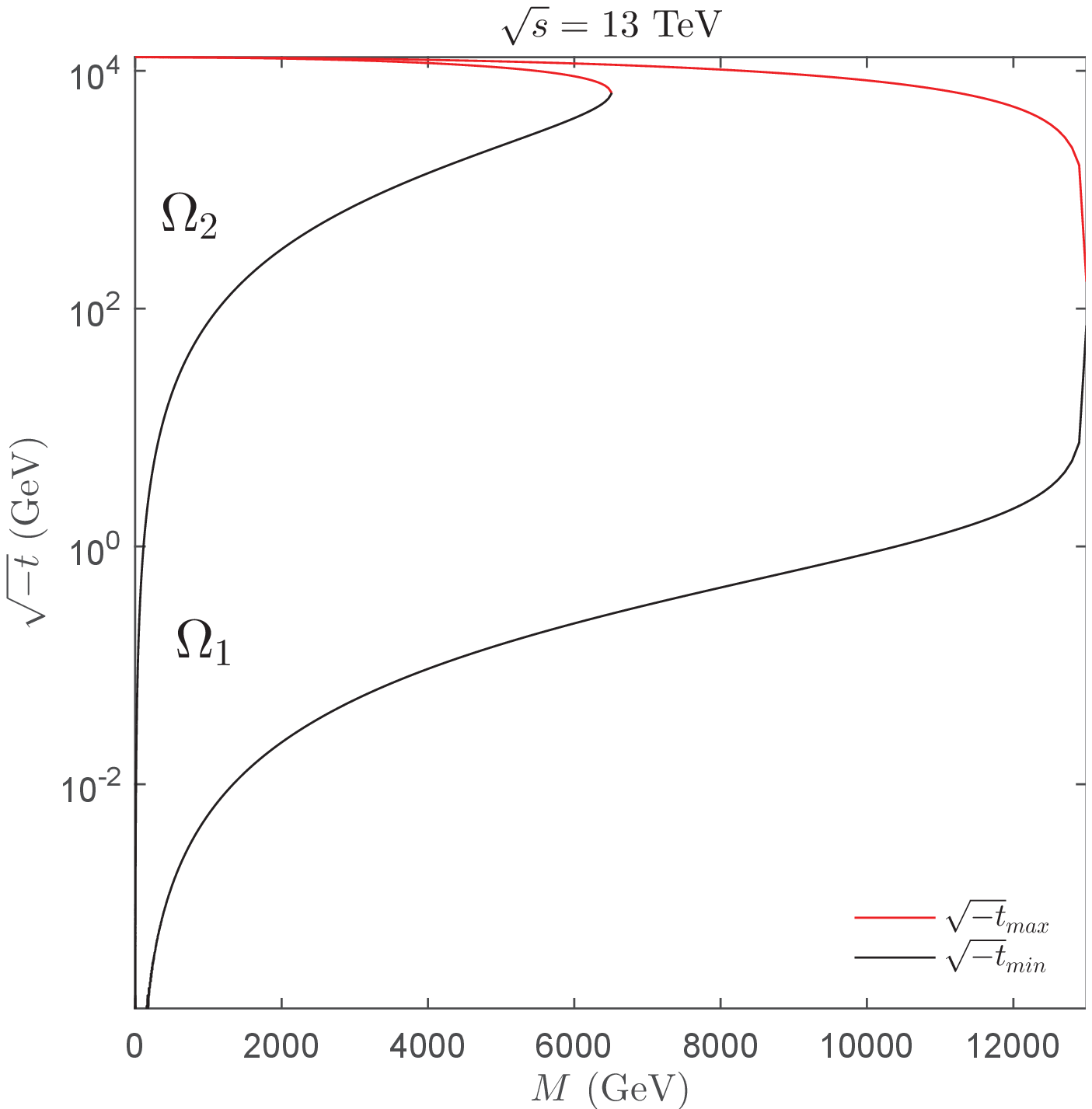}
  \caption{On left: The high mass "coherence condition" $\xi < 0.05 \sim x_F > 0.95 \sim \langle \Delta y \rangle > 3$ is only semi ad-hoc. On right: The phase space domain for $2 \rightarrow 2$ $t$-channel with variable invariant masses for 2 outgoing legs. $\Omega_1$ is for SD and $\Omega_2$ for DD with $M_X = M_Y$. The general case of DD $(M_X \neq M_Y)$ is between $\Omega_1$ and $\Omega_2$.}
  \label{fig: phase_space_13_TeV}
\end{figure}

When defining the measurement in terms of the invariant mass of the diffractive system $M_X$ (or $\xi = 1 - p_z'/p_z$ as $\xi \simeq M_X^2/s$), no direct geometrical fiducial definition exists. Unless the pseudorapidities and true rapidities are assumed to be approximately equal, $\eta \simeq y$, and the average kinematical relations are used for rapidity gaps in single diffraction: $\langle \Delta y \rangle_{SD} \simeq -\ln\left( M^2_X/s \right)$ and in double diffraction: $\langle \Delta y \rangle_{DD} \simeq -\ln\left( M_X^2M_Y^2/(m_p^2 s)\right), \langle y_0 \rangle \simeq \frac{1}{2}\ln\left( M_X^2/M_Y^2 \right)$. In practise, Monte Carlo chain definitions of diffractive mass acceptances are used, based on varying hadronization model assumptions of the diffractive systems. This includes, for example, non-trivial final state $p_\perp$ behaviour of soft processes, which is not understood from the first principles. The invariant masses of the SD or DD systems are not directly measured at the LHC, even if the leading proton longitudinal momentum is known, thereby allowing the use of 4-momentum conservation. Figure \ref{fig: phase_space_13_TeV} shows the phase space of single and double diffraction at the LHC. High mass double diffraction is seen to be kinematically constrained to be high-$|t|$ process, while single diffraction is not.

\section{EXTRACTING DIFFRACTIVE CROSS SECTIONS}

The following extraction of diffractive cross sections is based on \textit{probabilistic} inversion, density estimation or multidimensional fitting procedure. Concerning the earlier work by the authors, with real valued multivariate approaches, see \cite{mieskolainen2014bayesian, kuusela2010multivariate}. The term extraction is used here, because the measurement of inclusive diffractive cross sections is always model dependent at the LHC. The chosen framework aims at making the model dependence explicit and as transparent as possible. Each scattering process class $C_j$, such as single or double diffraction, is described in terms of a $d$-dimensional probability density or likelihood function $p(\mathbf{x}\,|\,\theta_j, \lambda)$ with $\mathbf{x} \in \mathbb{F}_2^d$. These densities are from MC simulations or possibly from simple parametrizations. That is, they give us likelihood of observing a binary vector final state $\mathbf{x}$ originating from the class $j$. In addition to likelihood functions, a priori probability distributions $p(\theta_j, \lambda\,|\,\alpha)$ are constructed for physics model parameters $\theta_j$ and detector simulation nuisance parameters $\lambda$, which can both be vectors. Variable $\alpha$ denotes a generic hyperparameter, a parameter of the parameter distribution. Many of these steps are often implicitly included in traditional large rapidity gap event analyses; here these are explicitly accounted for.

In a fully Bayesian treatment, the posteriori probability distributions $p(\theta_j, \lambda\,|\,\mathbf{x}, \alpha) \propto p(\mathbf{x}\,|\,\theta_j, \lambda) p(\theta_j, \lambda\,|\,\alpha)$ are obtained for each process class by first posing prior distributions for each process and their parameters, and then proceeding with Monte Carlo sampling of the parameter space. This can be a computationally heavy process depending on the number of free parameters, distribution shapes and correlations. Point estimates and credibility intervals are then obtained directly. For the process cross sections or fractions, a frequentist fitting via Maximum Marginal Likelihood is especially straightforward when the iterative Expectation-Maximization (EM) algorithm \cite{dempster1977maximum} is used. This approach maximizes the denominator or "evidence" in the Bayes formula with respect to the fractions: ${\textrm{arg}\,\textrm{max}}_{\{f_j\}} \prod_{i = 1}^N \sum_{j = 1}^{|C|} p(\mathbf{x}_i\,|\,C_j) f_j$, over a sample of $N$ events. After iterating, integrated process fractions are obtained as $\hat{f}_j \equiv \langle p(C_j|\mathbf{x}) \rangle_{\mathbf{x}}$ with $\sum_j \hat{f}_j = 1$. These can be scaled to physical cross sections with a van der Meer scan.

Multidimensional fitting allows also estimation of parameters such as the effective Pomeron intercept $\alpha_P(0) = 1 + \Delta_P$. The intercept is an interesting model parameter, not only because it controls asymptotic energy behaviour of cross sections, but also due to its controlling role of the differential mass distribution in the triple-Pomeron $(PPP)$ high mass limit and at $t \rightarrow 0$ as $d\sigma_{SD}/dM_X^2 \propto 1/(M_X^2)^{1+\Delta_P}$. To emphasize, the purity or background corrections are automatically taken into account here, because all major inelastic processes are simultaneously fitted. Since no explicit large rapidity gaps are required, diffractive cross sections can be extracted at the actual high mass limit.

\section{ALGEBRAIC REPRESENTATIONS}

All the different $r$-subspaces, $0 \leq r \leq d$, contained in our binary vector space of final states are encapsulated in the Grassmannian manifold $\mathbf{Gr}(r,d,\mathbb{F})$, in the object describing all possible $r$-dimensional subspaces in $\mathbb{F}^d$. This is a very rich object of algebraic geometry with variety of applications from coding theory to mathematical physics. The manifold is defined as $\mathbf{Gr}(r,d) = \{r \times d$ matrices with rank $r \}\setminus$row operations, with $\dim \mathbf{Gr}(r,d) = r(d-r)$. Using the Grasmannian subspaces allows to probe the Regge (vertex) factorization of type $ \frac{d^3\sigma_{DD}}{dM_X^2dM_Y^2dt} \sim \frac{d^2\sigma_ {SD}}{dM_X^2dt} \frac{d^2\sigma_{SD}}{dM_Y^2dt} / \frac{d\sigma_{EL}}{dt}$ by comparing specific partial cross sections $\sigma_k$ with a simple algorithm. This is experimentally feasible to do at the LHC, given the lacking differential measurement capabilities of $M_X^2$ and $t$. By comparing the different subspace combination rates, information about multi-Regge type of factorization is gained.

In practise, an algebraic representation of the binary data is needed. The most direct representation amounts to just counting the relative rates of $2^d$ different (or $2^d-1$ non-zero) final states, and then normalizing them to a probability vector $\mathbf{p}$. However, the components of \textit{ordinary moments} $m_k$ and the components of \textit{central moments} $\delta_k$, are also easily defined using the Kronecker (tensor) products as \cite{teugels1990some}
\begin{eqnarray}
&\mathbf{p} =
\left\langle
\left(\begin{array}{cc}
 1  &   -1 \\
 0  &   1 \\
\end{array}\right)^{\otimes \,d}
\left(\begin{array}{c}
 1  \\
 X_d  \\
\end{array}\right)
\otimes
\left(\begin{array}{c}
 1  \\
 X_{d-1}  \\
\end{array}\right)
\otimes
\cdots
\otimes
\left(\begin{array}{c}
 1  \\
 X_1  \\
\end{array}\right)
\right\rangle , \\
&m_k =
\left \langle \prod_{i = 1}^d X_i^{k_i} \right \rangle
=
\left\langle
\left(\begin{array}{c}
 1   \\
 X_d \\
\end{array}\right)
\otimes
\left(\begin{array}{c}
 1  \\
 X_{d-1} \\
\end{array}\right)
\otimes
\cdots
\otimes
\left(\begin{array}{c}
 1   \\
 X_1 \\
\end{array}\right)
\right\rangle_k , \\
&\delta_k = 
\left\langle \prod_{i = 1}^d (X_i - \langle X_i \rangle)^{k_i} \right\rangle
=
\left\langle
\left(\begin{array}{c}
 1   \\
 X_d - \langle X_d \rangle \\
\end{array}\right)
\otimes
\left(\begin{array}{c}
 1  \\
 X_{d-1} - \langle X_{d-1} \rangle \\
\end{array}\right)
\otimes
\cdots
\otimes
\left(\begin{array}{c}
 1   \\
 X_1 - \langle X_1 \rangle \\
\end{array}\right)
\right\rangle_k,
\end{eqnarray}
where $k = 1 + \sum_{i=1}^d k_i2^{i-1}$ (little endian binary expansion), $1 \leq k \leq 2^d$ and $k_i \in \mathbb{F}_2$ are used. The central moments describe the correlations $(\#\, 2^d-d-1)$ between any 2 or more subspaces (rapidity intervals). $X_i$ are the corresponding Bernoulli random variables.

The probability distributions of rapidity gaps $\Delta y$ for simplified Pomeron and Reggeon exchanges and random fluctuations are expected to be approximately \cite{khoze2010diffraction}
\begin{eqnarray}
&P_P(\Delta y) = c_P \exp(\Delta y(\alpha_P - 1)), \;\; \alpha_P \sim 1.08\,(\textrm{soft}) \dots 1.3\,(\textrm{hard}) \\
&P_R(\Delta y) = c_R \exp(\Delta y(2\alpha_R - \alpha_P - 1)),  \;\; \alpha_R \sim 1/2 \\
&P_F(\Delta y) = \frac{1}{\ell_F} \exp(-\Delta y/\ell_F),
\end{eqnarray}
which give the short range correlation length for a Reggeon with $\ell_R = -1/(2\alpha_R-\alpha_P-1) \sim 1$, the long range correlation length for a Pomeron with $\ell_P = 1/(\alpha_P - 1) \sim 10$ and for the fluctuations at Tevatron $\ell_F \sim 0.7 - 0.75$ \cite{khoze2010diffraction}. These examples motivate the present combinatorial construction which goes beyond the multidimensional fitting and extraction of cross sections presented earlier, and is now compatible with discussion about multigaps, gap destruction and rescattering and short/long range $y$-correlations. The algebraic representation chosen here is motivated by simple arguments that diffractive dissociation should represent the statistical dispersion $\langle F^2 \rangle - \langle F \rangle^2$ in the absorption probabilities of the diffractive eigenstates such as in \cite{kaidalov2001probabilities}, by the Good-Walker view.

The approach intrinsically includes the combinatorial AGK rules \cite{abramowsky1974agk}, that is, the density of particles over rapidity intervals with cut Pomerons. Basis for an experimental algorithm could be obtained, for example, by histogramming the multiplicity or charge information over $n$ different combinations. As an example of the AGK cutting rules: the total cross section for exchange of $\mu$ Pomerons, $\sigma_\mu^{tot}$, partial cross section $\sigma_\mu^{(\nu)}$ of a final state with a number of $\nu$ cut Pomerons and their ratio as given in \cite{levin1995pomeron}
\begin{equation}
\frac{\sigma_\mu^{(\nu)}}{\sigma_\mu^{tot}} = (-1)^{\mu - \nu} \frac{\mu!}{\nu!(\mu-\nu)!}(2^{\mu-1} - \delta_{0\nu}).
\end{equation}
Substituting for example $\mu = 2$ and $v = 0,1,2$, the usual alternating AGK factors of $1$, $-4$ and $2$, are obtained. In order to sum over $\mu$ one needs an explicit model, such as eikonal probabilities for the number of Pomerons being exchanged.


\section{COMBINATORIAL INVERSION}

The novel approach presented here also meets an interesting combinatorial challenge, which is the statistical inversion of "pileup" final states. These can be simultanenous proton-proton interactions at the LHC, but in principle any Poisson process which superimposes independent interactions, such as in the classic Miettinen-Pumplin model of wee partons \cite{miettinen1978diffraction}. The "direct model" equation is a convolution between Poisson and multinomial distributions as
\begin{eqnarray}
\label{eq: forwardmodel}
y_i = \frac{e^{-\mu}}{1-e^{-\mu}}\sum_{k = 1}^\infty \frac{\mu^k}{k!} \left\lbrace \sum_{ \Omega_{ik} } \frac{k!}{\prod_{j=1}^n x_j!}\prod_{j = 1}^n p_j^{x_j} \right\rbrace,
\end{eqnarray}
where $y_i$ is the pileup \textit{diluted} or \textit{enhanced} probability of observing $i$-th binary final state, $i = 1,\dots,2^d-1 = n$ and $\mu$ is the Poisson mean. The multinomial term in brackets and its values of $x_j\in \mathbb{N}$ are evaluated over all valid combinations generating the $i$-th final state $\mathbf{c}_i \in \mathbb{F}_2^d$ at Poisson order $k$ from the set of $n$-tuples $\Omega_{ik}$. Those which are allowed by partially ordered set (poset) combinatorics
\begin{equation}
\Omega_{ik} = \{ (x_1,\dots, x_j, \dots, x_n) \, | \, \bigvee_j x_j \mathbf{c}_j = \mathbf{c}_i \; \textrm{and} \sum_j x_j = k \},
\end{equation}
where the Boolean $\bigvee$ operator takes care of "summing" the binary vectors $\mathbf{c}_j$ of multiplicity $x_j$ and thus evaluating the pileup compositions.

\begin{figure}[ht]
\includegraphics[width=160pt]{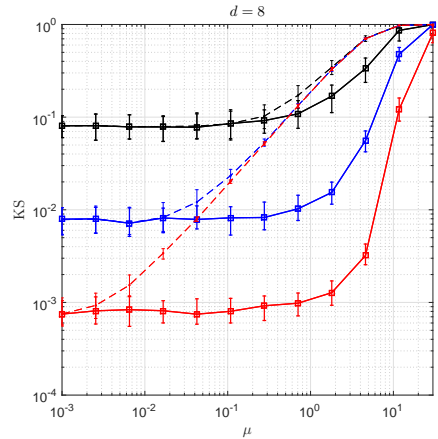}
  \caption{Inversion performance in solving $\mathbf{p}$ under Kolmogorov-Smirnov error (KS) as a function of Poisson $\mu$. Dashed lines without inversion. $N = 10^2$, (black), $10^4$, (blue), $10^6$ (red) events. Performance is fundamentally limited by $\sqrt{N}$ statistics and saturation at high $\mu$.}
\label{fig: KS_8}
\end{figure}

The basic idea is that the probabilities $\mathbf{y}$ are measured, and it is $\mathbf{p}$ to be solved by inverting Equation \ref{eq: forwardmodel}. An alternating sign solution similar to AGK rules can be obtained using the so-called principle of inclusion-exclusion (PIE) which is the M\"obius inversion for subsets in the combinatorial incidence algebra context \cite{rota1964foundations}, also utilized for example in mathematical physics in \cite{spector1990supersymmetry}. Exact details of the present combinatorial inversion will be discussed and presented elsewhere. Finally, a performance demonstration of the inversion algorithm is shown in Figure \ref{fig: KS_8}.

\section{ACKNOWLEDGEMENTS}
The organizers of Diffraction 2016 are gratefully acknowledged for an excellent conference.


\nocite{*} 
\bibliographystyle{aipnum-cp}%
\bibliography{sample}%

\end{document}